\documentclass{article} 

\usepackage{amsmath,amsfonts,bm}
\usepackage{todonotes} 
\usepackage{algorithm}
\usepackage{algorithmic}
\usepackage{amsmath}
\usepackage{amssymb}

\newcommand{\refSec}[1]{Section \ref{#1}}
\newcommand{\refFig}[1]{Fig. \ref{#1}}

\def\eqref#1{equation~\ref{#1}}

\def\1{\bm{1}}

\DeclareMathAlphabet{\mathsfit}{\encodingdefault}{\sfdefault}{m}{sl}
\SetMathAlphabet{\mathsfit}{bold}{\encodingdefault}{\sfdefault}{bx}{n}

\usepackage{wrapfig}
\usepackage{graphicx}
\usepackage{booktabs} 
\usepackage{multirow} 
\usepackage{diagbox} 
\usepackage{multicol} 

\usepackage{amsmath}     
\usepackage{amssymb}     
\usepackage{iclr2026_conference,times}

\usepackage{hyperref}
\usepackage{url}

\newcommand{\chg}[1]{{\color[rgb]{0.0,0.0,0.0}#1}}

\iclrfinalcopy 

\title{DISK: Differentiable Sparse Kernel Complex for Efficient Spatially-Variant Convolution}

\author{Zhizhen Wu\thanks{Equal contribution.}\quad 
Zhe Cao\footnotemark[1]\quad
Yuchi Huo\thanks{Corresponding author.} \\[0.3em]
State Key Lab of CAD\&CG, Zhejiang University, China \\[0.3em]
\texttt{zhizhenwu@zju.edu.cn, caozhe022@qq.com, huo.yuchi.sc@gmail.com}
}

\begin{document}

\maketitle
\lhead{Published as a conference paper at ICLR 2026}
\begin{figure}[h]
    \centering
    \includegraphics[width=\columnwidth]{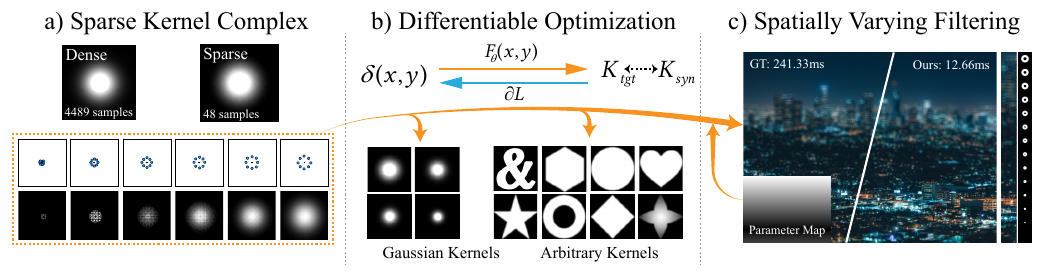}
    \caption{\textbf{An overview of our method.} We represent a dense filter as a \textit{Sparse Kernel Complex}, a sequence of sparse layers whose parameters $\Theta$ are learned via differentiable optimization. We apply our filter $F_{\Theta}$ to an impulse $\delta$ to yield a synthesized kernel $K_{syn}$, and minimize a loss $\mathcal{L}$ against the target $K_{tgt}$ to learn arbitrary shapes. These optimized kernels form a basis for high-performance spatially varying filtering, achieving quality close to ground truth with up to a 20$\times$ speedup.}
    \label{fig:overview}
\end{figure}
\begin{abstract}


Image convolution with complex kernels is common in photography, scientific imaging, and animation, but dense convolution is too expensive for resource-limited devices. Existing approximations, such as simulated annealing and low-rank decompositions, are either slow or struggle with non-convex kernels.
We present a differentiable kernel decomposition framework that represents a spatially variant dense kernel with a small set of sparse samples, assuming the target dense kernel is known for both optimization and filtering. Our method provides (i) end-to-end differentiable sparse-kernel optimization, (ii) shape-aware initialization for non-convex kernels, and (iii) kernel-space interpolation for efficient, multi-dimensional spatially varying filtering without retraining or added runtime cost.
Across Gaussian and non-convex kernels, our method achieves higher fidelity than simulated annealing and lower cost than low-rank decomposition. It is practical for mobile imaging and real-time rendering, and integrates cleanly into learning pipelines.

\end{abstract}
\section{Introduction}

From rendering realistic depth-of-field effects~\citep{sakurikarepsilon,wu2022dof} in computational photography to modeling the intricate point spread functions~\citep{liu2022deep,shajkofci2020spatially} of optical systems, the ability to apply large, complex convolution kernels is a fundamental building block in modern vision and graphics computing systems. This creates a fundamental tension: while larger, more intricate kernels enable higher-fidelity results, their quadratic computational cost renders direct implementation impractical for interactive applications on devices ranging from mobile phones to high-end GPUs. 

To bridge this gap, many works have focused on approximation strategies. For specific cases like Gaussian blur, elegant constant-time solutions~\citep{zing2010extended,kovesi2010fast} exploit the filter’s analytic structure. However, such specialized methods are not applicable to the arbitrary, often non-convex kernels needed for advanced effects. More general approaches, such as low-rank matrix decomposition~\citep{mcgraw2015fast}, support arbitrary kernels but typically reduce the computation to a sequence of smaller dense convolutions, which limits sparsity and caps efficiency gains.

A more direct and efficient approach~\citep{schuster2020high} is to approximate a dense kernel with a truly sparse one, drastically reducing the number of required computations. It relies on heuristic search via parallel simulated annealing to identify sparse sampling patterns for arbitrary kernels. Despite its generality, it requires many iterations and still misses high-fidelity solutions due to the non-convex optimization landscape. This motivates a more principled and efficient way to obtain high-quality sparse kernel representations.

In this work, we introduce a differentiable kernel decomposition framework to address this challenge. Like prior approximation methods, our formulation assumes access to the dense target kernel throughout optimization and filtering. Unlike traditional low-rank methods, we optimize a sequence of natively sparse kernels, yielding an efficient representation for runtime inference. We cast the decomposition as an end-to-end optimization problem and apply gradient-based methods instead of heuristic search such as simulated annealing, enabling more reliable convergence to high-fidelity solutions in far fewer iterations. To improve stability for non-convex target kernels, we further propose a two-part initialization strategy: structure-aware sampling to capture fine shape details and a deterministic radial initialization to stabilize and accelerate convergence.

Beyond single-kernel approximation, our framework supports efficient, multidimensional spatially varying (SV) filtering. The main challenge is the cost of generating a unique kernel per pixel, which often creates a significant performance bottleneck. We address this with filter-space interpolation: we precompute an optimized basis of sparse filters spanning the target effect range, then synthesize per-pixel kernels at runtime by interpolating this compact basis. This strategy reduces per-pixel synthesis to a minimal set of multiply-add operations, decoupling kernel-generation cost from image resolution and enabling sophisticated SV effects with negligible overhead.

Our contributions are as follows:
\begin{itemize}
\item A differentiable framework for decomposing a dense, arbitrary kernel into a sequence of optimized sparse layers, enabling efficient, high-fidelity approximation.
\item An initialization scheme, combining a general radial strategy for stable convergence with a sparse sampling method for capturing non-convex kernels.
\item A filter-space interpolation method for high-performance, spatially-varying filtering that decouples kernel synthesis cost from image resolution.
\end{itemize}
\section{Related Work}

\subsection{High-Performance Kernel}
Given that Gaussian blur is computationally expensive, numerous methods have been proposed to accelerate it. Common examples include $O(1)$ approximations such as the Extended Binomial Filter~\citep{zing2010extended} and Summed-Area Table-based methods~\citep{kovesi2010fast}. However, their reliance on precomputation or inherently sequential steps makes them ill-suited to the massively parallel design of modern GPUs.

A better match for real-time rendering is Kawase blur~\citep{kawase2003frame}, a multi-pass filter that uses only four texture samples per pass. As an extension, Dual Filtering~\citep{Martin2015dual} introduces downsampling followed by upsampling, reducing bandwidth and computation by operating on lower-resolution textures. However, such pipelines trade regular downsampling for extra reconstruction work (often additional convolution) and may lose fine detail, whereas we stay at a single resolution and reduce cost via flexible sparse sampling while matching the target kernel response. A further practical limitation is the lack of a systematic mapping from a target Gaussian strength (e.g., a given $\sigma$) to the corresponding Kawase or Dual Filtering parameters; our work addresses this gap.

\subsection{Spatially-Variant Filtering}
A substantial body of work learns spatially varying, per-pixel convolution kernels, with applications to video prediction, frame interpolation, denoising, and deblurring~\citep{jia2016dynamic, niklaus2017video, mildenhall2018burst, zhou2019spatio, zhou2021decoupled}. Unlike approaches that directly predict a dense per-pixel kernel map, our method decouples filter generation from spatial resolution. Specifically, we learn a compact lookup table (LUT) that parameterizes a continuous family of filters, enabling flexible and efficient spatially-variant filtering.
Spatiotemporal Variance-Guided Filtering~\citep{schied2017spatiotemporal} forms per-pixel filter mixtures guided by estimated spatial and temporal variance. Differently, we condition filter generation on an input per-pixel blur-intensity map. This allows us to synthesize filters that match any target spatially varying blur, without intermediate variance estimation or other content statistics.

\subsection{Kernel Approximation and Decomposition}
Inspired by Kawase blur~\citep{kawase2003frame}, High-Performance Image Filters~\citep{schuster2020high} uses parallel tempering to optimize sparse sampling patterns, but its optimization stability can be limited by sensitivity to many hyperparameters compared with our gradient-based formulation.
Kernel decomposition is also used to reduce cost in other settings: 2D kernels into pairs of 1D kernels for video interpolation~\citep{niklaus2017video}; 3D spatiotemporal kernels into spatial and temporal atoms (STDCF)~\citep{schied2017spatiotemporal}; standard convolution into depthwise and $1\times1$ pointwise operations~\citep{howard2017mobilenets, chollet2017xception, ramadhani2024improving}; dynamic weights into static bases plus residuals~\citep{li2021revisiting, li2024unet}; and related decompositions for graph-transformer attention (KDLGT)~\citep{wu2023kdlgt} and large depthwise kernels (LKD-Net)~\citep{luo2023lkd}.

Despite their success, these methods are largely limited to structured factorizations or heuristic optimization. Our framework addresses the missing capability: a differentiable, general solution for sparse approximation of arbitrary (including non-convex) kernels, together with an efficient mechanism for spatially varying kernel synthesis.


\section{Preliminary}
\subsection{Kernel-Based Filtering}
Kernel-based filtering is fundamental to many image processing tasks. This process takes an input image $I_{in}$ and computes each pixel's value for the output image $I_{out}$ as a weighted average of its local neighbors within $I_{in}$. Formally, this operation is expressed as a 2D convolution, defined as:
\begin{equation}
\label{eq:convolution}
    I_{out}[x, y] = (I_{in} * K)[x, y] = \sum_{i=-k}^{k} \sum_{j=-k}^{k} I_{in}[x+i, y+j]\cdot K[i, j],
\end{equation}
where the matrix $K$ is the $M \times M$ convolution with kernel size $M\in \mathbb{R}^{+}$, whose elements $K[i, j]$ are weights that determine the contribution of each neighboring pixel to the final filtered value. 
\subsection{Filter Representation}
The dense matrix representation for the kernel $K$ in Eq. (1) is straightforward. However, its $O(M^2)$ computational cost presents a significant bottleneck. This is especially true for filters with a large spatial support, such as a Gaussian blur with a large $\sigma$, where the cost becomes prohibitively expensive for real-time applications that demand high frame rates.

Our key insight is to approximate this expensive operation by structuring the filter as a sequence of lightweight convolutional layers, where the output of one layer serves as the input for the subsequent one. Each layer applies a highly efficient sparse kernel, $K_{sparse}$, which we define by a small collection of $N$ samples with offset-weight pairs:
\begin{equation}
    K_{sparse} = \{(\mathbf{o}_i, w_i)\}_{i=1}^N,
\end{equation}
where $\mathbf{o}_i \in \mathbb{R}^2$ is the spatial offset and $w_i$ is its corresponding weight.

The complete operation, consisting of $L$ such layers with kernels ($K_1, K_2, ..., K_L$), can be expressed as a nested convolution:
\begin{equation}
    I_{out} = (...((I_{in} * K_1) * K_2) * ... * K_L).
\end{equation}
This multi-layer filter reduces the cost to $O(\sum_{l=1}^{L} N_l)$ per pixel. Since this sum is far smaller than the number of weights in the target dense kernel ($ \sum N_l \ll M^2$), the approach offers a dramatic speedup.

\section{Methodology}

\subsection{Differentiable Multi-Layer Kernel Complex}
\paragraph{Overview}
Sparse filters offer a computationally efficient alternative to dense kernels; however, they often fail to capture the intricate structure of large, complex filters. The core challenge lies in determining the optimal parameters—the spatial offsets and weights—for a sequence of sparse kernels to accurately reconstruct a target. Manually designing these parameters or using traditional, non-differentiable methods is a formidable task.

To overcome this, our key contribution is to frame the decomposition as a differentiable optimization problem. This enables the simultaneous end-to-end learning of all sparse kernel parameters across all layers. We define the complete set of these learnable parameters as $\Theta =\{(\mathbf{o}_{l,i}, w_{l, i})\}_{l=1, i=1}^{L, N_l}$, which includes the offsets and weights for $N_l$ samples in each of the $L$ layers.

Our goal is to find the optimal parameters $\Theta^*$ by minimizing a loss function $\mathcal{L}$ that measures the discrepancy between our approximation and the target kernel:

\begin{equation}
\label{eq:optim-target}
\begin{aligned}
    \Theta^* = \arg\min_{\Theta} \mathcal{L}(K_{target}, F_{approx}(\Theta)) ,\\
    F_{approx}(\Theta) = K_{s, 1}* K_{s, 2}* ...* K_{s, L},
\end{aligned}
\end{equation}

where $K_{target}$ is the desired dense filter and $F_{approx}(\Theta)$ is the composite kernel formed by the convolution of the learned sparse kernels.

\paragraph{Learnable Parameter}
Our optimization strategy treats the offsets and weights of each sample as independent, learnable parameters. Specifically, for each layer $l$ and for each of the $N_l$ sampling points within it, we simultaneously optimize both the 2D offset vector $\mathbf{o}_{l, i}$ and its corresponding scalar weight $w_{l, i}$.

The complete set of learnable parameters for the entire model, denoted by $\Theta$, is therefore the collection of all such offset-weight pairs:
\begin{equation}
    \Theta = \bigcup_{l=1}^{L} \{(\mathbf{o}_{l,j}, w_{l,j})\}_{j=1}^{N_l}.
\end{equation}


\paragraph{Initialization}
A well-designed parameter initialization is crucial for stable optimization convergence. Heuristic methods, such as Kawase~\citep{kawase2003frame} and Dual Filtering~\citep{Martin2015dual}, have fixed schemes tailored to specific filter types; however, a general approach is required for arbitrary target kernels of different sizes.

To address this, we propose a radial initialization strategy. The core idea is to initialize the sampling points in each layer to be uniformly distributed on the circumference of a circle, with the radius of this circle increasing linearly with the layer index. This progressive expansion ensures that the effective receptive field of the composite kernel grows with each subsequent layer, making the initial configuration capable of spanning a large-area target kernel from the outset.  The radius for layer $l$, denoted $r_l$, is governed by a step size $\Delta_r$ derived from the target kernel's spatial extent and the total number of layers $L$ (see Appendix for derivation). The corresponding weights in each layer are initialized uniformly.

This initialization is formally defined as:

\begin{equation}
\begin{aligned}
    r_l &= l \cdot \Delta_r &&\text{for } l = 1, \dots, L, \\
    \mathbf{o}_{l,i} &= \left( r_l \cos\left(\frac{2\pi i}{N_l}\right), r_l \sin\left(\frac{2\pi i}{N_l}\right) \right) &&\text{for } i = 1, \dots, N_l, \\
    w_{l,i} &= \frac{1}{N_l}.
\end{aligned}
\end{equation}


\subsection{Sparse Sampling of Arbitrary Kernel}
A common way to initialize filter offsets is by sampling random positions within a local neighborhood. While this approach is general, it often traps the optimization in poor local optima, especially for kernels with complex or non-convex shapes.

Our method decomposes a dense kernel into a series of sparse ones. The first of these, $K_{s,1}$~(Eq.~\ref{eq:optim-target}), has the greatest influence on the final filtered output, so its initialization is critical. A simple improvement over purely random sampling is to confine samples to the minimal bounding box of the kernel's non-zero pixels. This ensures most samples fall near the target shape, but it is still inefficient for non-convex kernels, whose bounding boxes can contain large empty regions.

To overcome this limitation, we propose a more sophisticated initialization strategy leveraging rejection sampling. Instead of drawing samples from the kernel's bounding box, our method samples directly from the support of the kernel, i.e., its non-zero locations. We first quantify the effective sampling area, denoted by $S$, as the count of these non-zero pixels. A sampling radius $r$ is subsequently derived based on the desired number of samples, $N_s$:
\begin{equation}
r = \sqrt{\frac{S}{N_s \cdot \pi}}.
\end{equation}
The detailed procedure is provided in the appendix. This initialization sets the first sparse kernel’s offsets to closely match the target shape. By restricting samples to relevant regions, it avoids vanishing gradients and reduces the risk of converging to poor local optima.

\subsection{Spatially Varying Filtering}
Next, we propose a decomposition method for spatially varying filtering.

Spatially varying filtering generalizes convolution by applying a unique filter at each pixel $(x,y)$. The filter's properties---such as its blur radius, orientation, or shape---are determined by a corresponding value $P(x,y)$ from a parameter map. The core challenge lies in efficiently synthesizing and applying these unique per-pixel kernels.

Conventional spatially varying filtering is often impractical. Generating dense kernels on the fly~\citep{wang2023implicit} is too slow, while precomputing them~\citep{kovesi2010fast} incurs prohibitive memory costs; both are a poorly suited to modern parallel hardware. Faster alternatives~\citep{leimkuhler2018laplacian} restrict filters to simple analytic forms (e.g., Gaussians), but this sacrifices the expressiveness needed for complex, non-convex point spread functions (PSFs).

We observe that, in prior work, the cost of generating or storing spatially varying kernels grows linearly with image resolution. To remove this bottleneck while retaining expressive, sparsely optimized kernels, we introduce \textit{Filter-Space Interpolation}, which decouples kernel-generation complexity from image size.

Our spatially varying filtering is built on an ordered set of $M$ basis sparse filters, which discretely sample a continuous, one-dimensional space $\mathcal{F}$ of filters. Each basis filter, $f_k$, corresponds to a scalar parameter $p_k$ (with $p_1 < p_2 < \dots < p_M$) and consists of a unique set of $N$ sampling offsets and weights. This design allows our basis to represent a wide range of filter behaviors across the parameter space, from applying arbitrary linear transformations to a kernel to simply varying the standard deviation ($\sigma$) of a Gaussian. We define the basis as:
\begin{equation}
    \mathcal{F} = \left\{ f_k(p_k) \mid k = 1, \dots, M \right\}, \quad \text{where} \quad f_k = \left\{ (\mathbf{o}_{ki}, w_{ki}) \right\}_{i=1}^{N}
    \label{eq:basis-filters}
\end{equation}
We divide the approach into an offline pre-computation stage and a runtime inference stage. 
In the offline stage, we optimize each basis filter $f_k$ individually to represent the ideal filter effect at its parameter value $p_k$.

At runtime, we synthesize a unique sparse filter for each pixel $(x,y)$, which is guided by a per-pixel parameter map, $P$. From the parameter value at each coordinate, $P(x,y)$, we determine a corresponding vector of $M$ interpolation weights, $\bm{\alpha}(x,y) = (\alpha_1, \dots, \alpha_M)$. These weights specify how to blend a compact set of basis filters, $\{f_k\}_{k=1}^M$, to reconstruct the final filter instance.

The final sparse filter for a given pixel, $f(x,y)$, is synthesized as a direct convex combination of the basis filters:
\begin{equation}
    f(x, y) = \sum_{k=1}^{M} \alpha_k(x, y) \cdot f_k,
    \label{eq:filter_interpolation}
\end{equation}
subject to the constraint that $\sum_{k=1}^{M} \alpha_k(x, y) = 1$ and $\alpha_k(x, y) \ge 0$. 

\chg{By directly interpolating basis-filter offsets and weights, we sidestep the costly on-the-fly generation of kernels from analytical functions.} This reduces the computational overhead of spatially varying kernel synthesis to a minimal set of parallelizable multiply-add operations. Furthermore, the interpolatable nature of our basis filters makes the entire set highly compressible, allowing us to significantly reduce the memory footprint required to achieve a wide range of expressive effects while offering flexible control over the quality-performance trade-off.

\subsection{Implementation Details}
\paragraph{Training Process}
To ensure our learned filter parameters are generalized and not overfit to a specific dataset, we adopt an image-agnostic optimization strategy. We leverage a core principle of Linear Shift-Invariant (LSI) systems\chg{~\citep{goodman2005introduction}}: a filter is fully characterized by its impulse response.

First, we synthesize the effective kernel of our multi-pass filter, $F_\theta$, by applying it to a discrete Dirac delta function, $\delta$. The resulting output is the synthesized impulse response, $K_{\text{syn}}$. The impulse $\delta$ is an image with a single non-zero pixel at its center coordinate $\mathbf{c}$:
\begin{equation}
    K_{\text{syn}} = F_\theta(\delta), \quad \text{where} \quad \delta[\mathbf{n}] =
    \begin{cases}
        1 & \text{if } \mathbf{n} = \mathbf{c} \\
        0 & \text{otherwise.}
    \end{cases}
\end{equation}
Here, $\theta$ represents the learnable parameters of our filter and $\mathbf{n}$ denotes the discrete pixel coordinates.

\paragraph{Loss Design}
Second, we define our loss function, $\mathcal{L}$, as the Charbonnier L1 loss $\mathcal{C}$~\citep{charbonnier1994two} between the synthesized kernel $K_{\text{syn}}$ and a target kernel $K_{\text{tgt}}$:
\begin{equation}
    \mathcal{L} = \mathcal{C}(K_{\text{syn}}, K_{\text{tgt}}). 
\end{equation}
This impulse-response-based supervision allows us to "collapse" the entire multi-layer filtering sequence into a single, equivalent kernel for direct and precise approximation of the target.



\section{Experiments}
In this section, we conduct a series of experiments to evaluate our differentiable kernel decomposition framework thoroughly. We first describe the experiment details and evaluation protocol in~\refSec{sec:exp-setup}. Next, in~\refSec{sec:exp-single-kernel}, we assess our method's ability to approximate single, complex kernels, comparing it against state-of-the-art techniques. \chg{We extend this analysis to the more challenging task of spatially varying filtering in~\refSec{sec:exp-svk}.  To validate our specific design choices, we present a series of ablation studies in~\refSec{sec:exp-ablation}.}
\begin{figure}[t]
    \centering
    \includegraphics[width=0.497\linewidth]{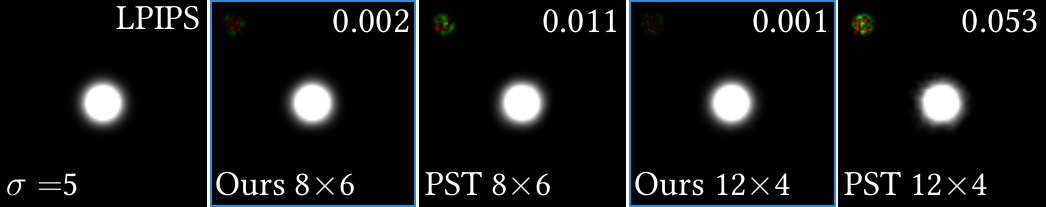}
    \hfill
    \includegraphics[width=0.497\linewidth]{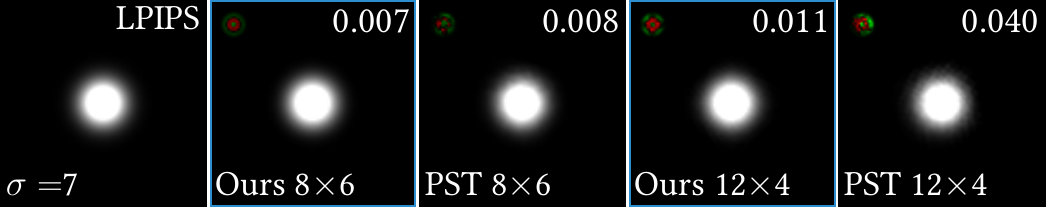}
    \includegraphics[width=0.497\linewidth]{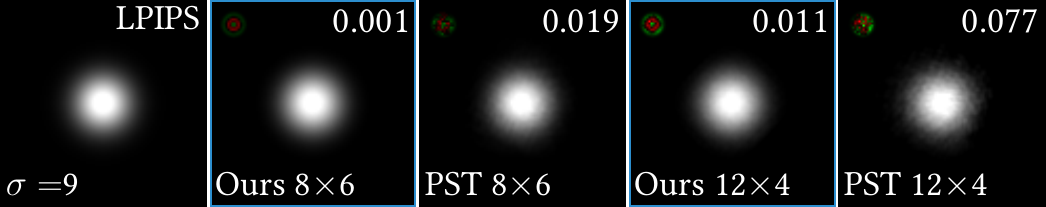}
    \hfill
    \includegraphics[width=0.497\linewidth]{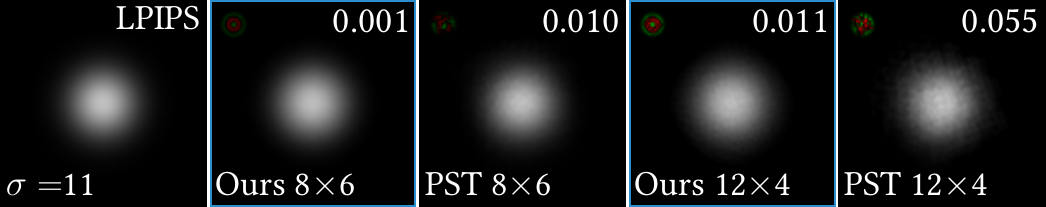}
    \caption{\textbf{Comparison of Gaussian kernel approximation with varying $\sigma$.} We compare our method against PST using two sparse configurations (8 layers × 6 samples and 12 layers × 4 samples). LPIPS scores appear in the top-right corner (lower is better). The top-left inset visualizes the error map, with positive errors shown in red and negative errors in green.}
    \label{fig:gs-comparison}
\end{figure}
\subsection{Setup}
\label{sec:exp-setup}
\paragraph{Baselines.}
We compare our method against several baselines. For both single kernel and spatially varying filtering, we include a \chg{low-rank decomposition} (LowRank)~\citep{mcgraw2015fast} and the optimization-based method of Parallel Tempering (PST)~\citep{schuster2020high}. 
\paragraph{Datasets and Kernels.}
To evaluate the versatility of our method, we use a diverse set of target kernels and images. This set includes standard analytical shapes, such as Gaussian kernels (with $\sigma$ values from 5 to 11). \chg{To assess performance on more complex targets, we additionally use a suite of arbitrary kernels comprising simple geometric primitives (disks, rings), regular polygons (4-sided and 6-sided), non-convex shapes (a heart, a four-pointed star, and an ampersand symbol), more complex shapes (animal silhouettes), and optical PSFs (coma and spherical aberration).
For the spatially varying filtering experiments, we use five high-resolution photographs selected to represent realistic scenarios with complex textures and both 1D and 2D spatial variations.}
\paragraph{Implementation and Evaluation Metric.}
We implement our methods in PyTorch and perform all optimization on a single GPU with 24 GB of memory, offering computational power comparable to an NVIDIA RTX 4090. For all configurations of kernels and layers, we use the same Adam optimizer with a learning rate linearly decayed from $1\times 10^{-3}$ to $1\times 10^{-4}$. We use 1000 optimization steps per kernel for our method. For comparison, we run the PST algorithm for 10,000 iterations with 10 parallel candidates, for a total of 100,000 optimization steps. For the LowRank method, we utilize decompositions with ranks 1,2 and 3, chosen to maintain a comparable number of samplings.

\begin{wrapfigure}{r}{0.45\textwidth} 
    \centering
    \vspace{8pt}
    \includegraphics[width=0.95\linewidth]{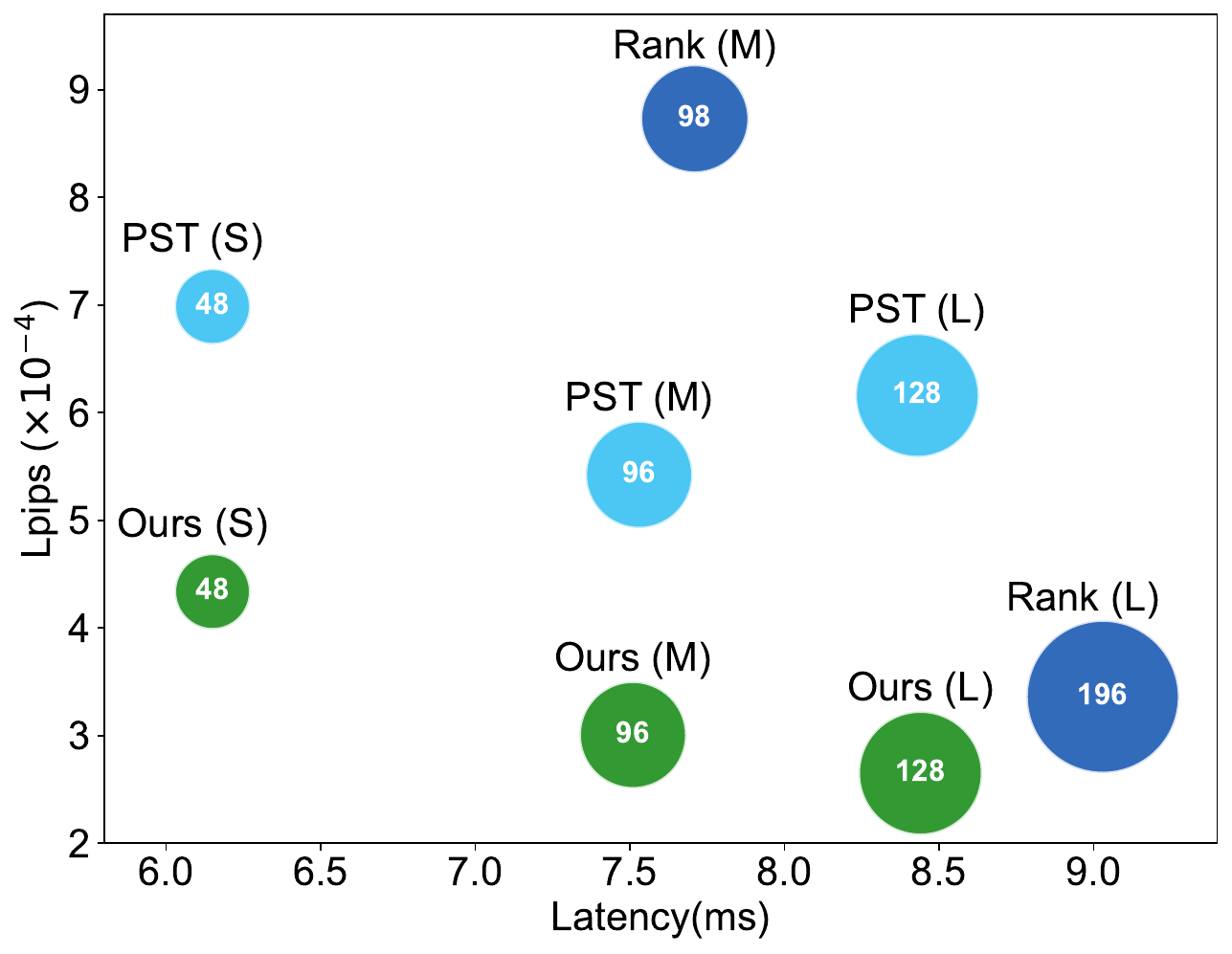} 
    \caption{\small \textbf{Speed, accuracy, and samples comparison.}  The figure plots quality against latency (lower is better for both). The size of each bubble represents the total sample count.}
    \vspace{-16pt}
    \label{fig:npass-plot}
\end{wrapfigure}

For runtime analysis, we benchmark our approach on a representative mobile device equipped with a Qualcomm Snapdragon 8 Gen 3 SoC, and report latency in milliseconds (ms). We evaluate both numerical fidelity and perceptual similarity using Peak Signal-to-Noise Ratio (PSNR), Learned Perceptual Image Patch Similarity (LPIPS)~\citep{zhang2018unreasonable}, and FLIP-LDR~\citep{andersson2020flip}. Higher values indicate better performance for PSNR, and lower values are better for LPIPS and FLIP-LDR.
\subsection{Single Kernel}
\label{sec:exp-single-kernel}
\refFig{fig:npass-plot} shows that our method consistently achieves a superior balance between reconstruction quality and inference speed compared to all other approaches.  For our method and PST, the 'S', 'M', and 'L' correspond to total sample counts of 48 (12$\times$4), 96 (24$\times$4), and 128 (32$\times$4), respectively. The LowRank's 'M' and 'L' use 98 (49$\times$2) and 196 (49$\times$4) parameters.

Next, we present a comparison of Gaussian kernel approximation with varying standard deviations $\sigma$ in ~\refFig{fig:gs-comparison}.
In a 6-layer, 8-sample (8$\times$6) configuration, our method achieves high-fidelity results with low perceptual error, whereas PST exhibits visible noise and artifacts. This performance gap widens in a sparser 12$\times$4 setup. As $\sigma$ increases, PST's approximation degrades severely, while our result remains visually coherent and maintains a substantially lower LPIPS error.
These results show that our gradient-based optimization yields more accurate approximations than PST, consistently producing stable solutions even in challenging sparse configurations.
\begin{figure}[t]
    \centering
    \includegraphics[width=\columnwidth]{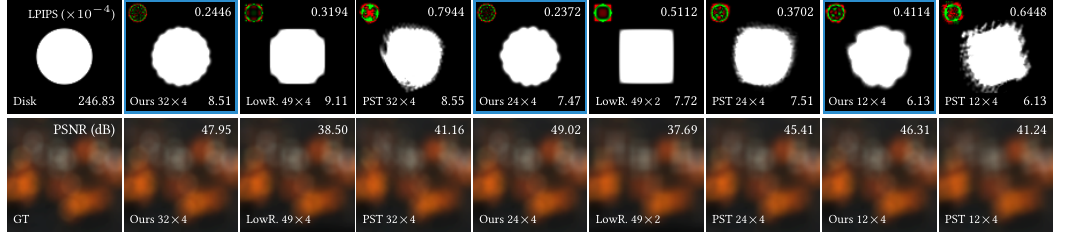}
    \includegraphics[width=\columnwidth]{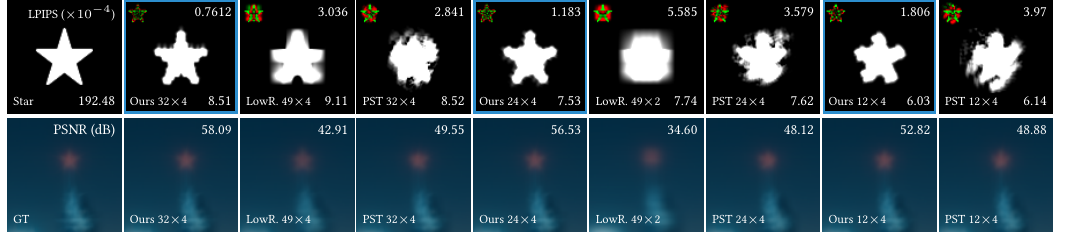}
    \includegraphics[width=\columnwidth]{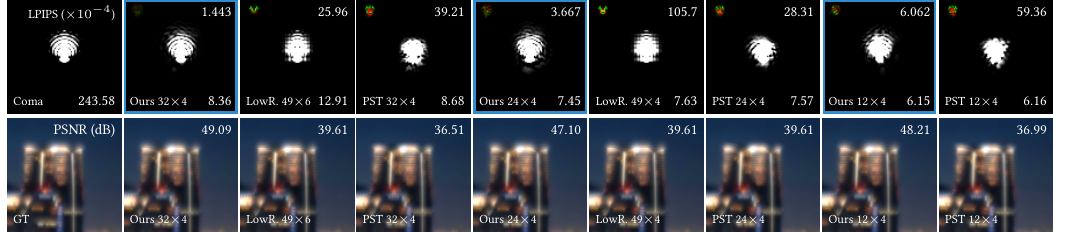}
    \caption{\textbf{Comparison of Single kernel approximation.} Compared to baselines, SVD-based decomposition (LowR.)~\citep{mcgraw2015fast} and Parallel Simulated Tempering (PST)~\citep{schuster2020high}, our approach (blue) better preserves sharp features on non-convex targets, resulting in lower LPIPS scores (lower is better). The top-left inset visualizes the error map, with positive errors shown in red and negative errors in green.}
    \label{fig:single-kernel-comparison}
\end{figure} 

\chg{Our method’s accuracy extends beyond Gaussian kernels to the more general case of arbitrary single-kernel filters, as shown in~\refFig{fig:single-kernel-comparison}.} Our method preserves structure across both simple and complex shapes, while LowRank introduces blocky artifacts and PST yields noisy results at low sample counts. Quantitatively, our method achieves the lowest LPIPS across all tests, often by a large margin. It is also much more efficient, requiring only $1/100$ the iterations of PST.

\subsection{Spatially Varying Kernel}
\label{sec:exp-svk}
\begin{figure}[t]
    \centering
    \includegraphics[width=\linewidth]{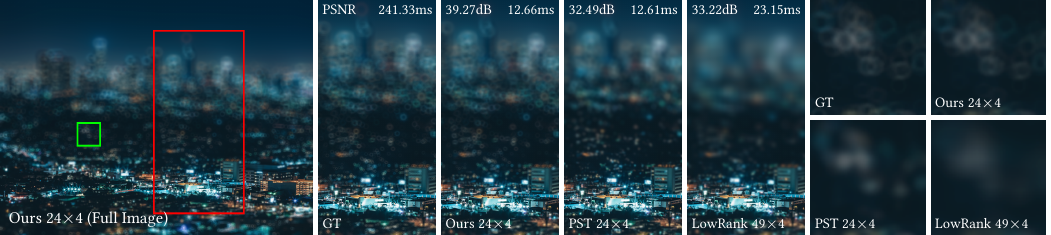}
    \includegraphics[width=\linewidth]{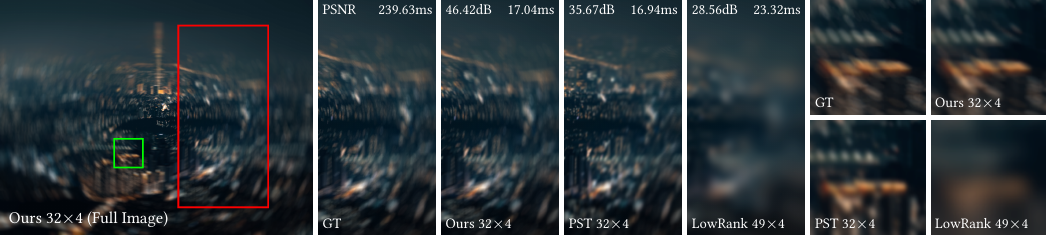}
    \includegraphics[width=\linewidth]{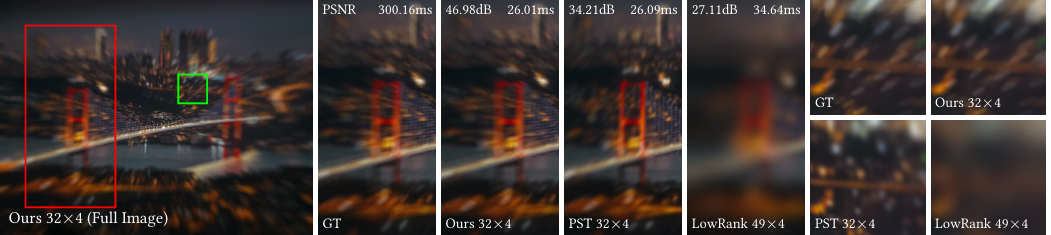}
\caption{\textbf{Visual comparison of diverse spatially varying (SV) effects.}
\chg{
We evaluate three SV configurations: 1D tilt-shift blur (top), 2D rotational blur (middle), and 2D radial motion blur (bottom). We compare our method against Parallel Simulated Tempering (PST) and Low-Rank Decomposition (LowRank).
}
}
    \label{fig:sv-comparison}
\end{figure}
\chg{We present three spatially varying filtering examples in~\refFig{fig:sv-comparison}. The first is a 1D spatially varying blur that uses a pseudo-depth map to simulate a tilt-shift camera effect. The other two are 2D anisotropic effects: a rotational bokeh blur and a radial motion blur, both controlled by two parameters—blur intensity and local blur angle.} 

Our results are visually indistinguishable from the ground truth. As highlighted in the red and green insets, our method reproduces the complex structure of the GT kernels. In contrast, PST introduces noise and LowRank oversmooths, and neither recovers the correct kernel shape, while directly applying GT kernels is prohibitively slow. Quantitatively, our method achieves the highest PSNR among all methods while maintaining real-time performance.

This performance difference stems from how well each method's base kernels handle filter-space interpolation. While all approaches use interpolation to generate the varying filter parameters, our optimization-based kernels are better conditioned for this process and appear to vary more linearly. 
Consequently, they interpolate smoothly to form sharp, complex patterns. PST's kernels, however, suffer from poor optimization quality, and interpolating between them simply produces more noise. Similarly, interpolating the basis kernels from LowRank's decomposition causes them to average into indistinct blurs rather than preserving the target structure.

\subsection{Ablations}
\label{sec:exp-ablation}
\begin{figure}[t]
    \centering
    \begin{minipage}{0.495\columnwidth} 
        \centering
        \includegraphics[width=\linewidth]{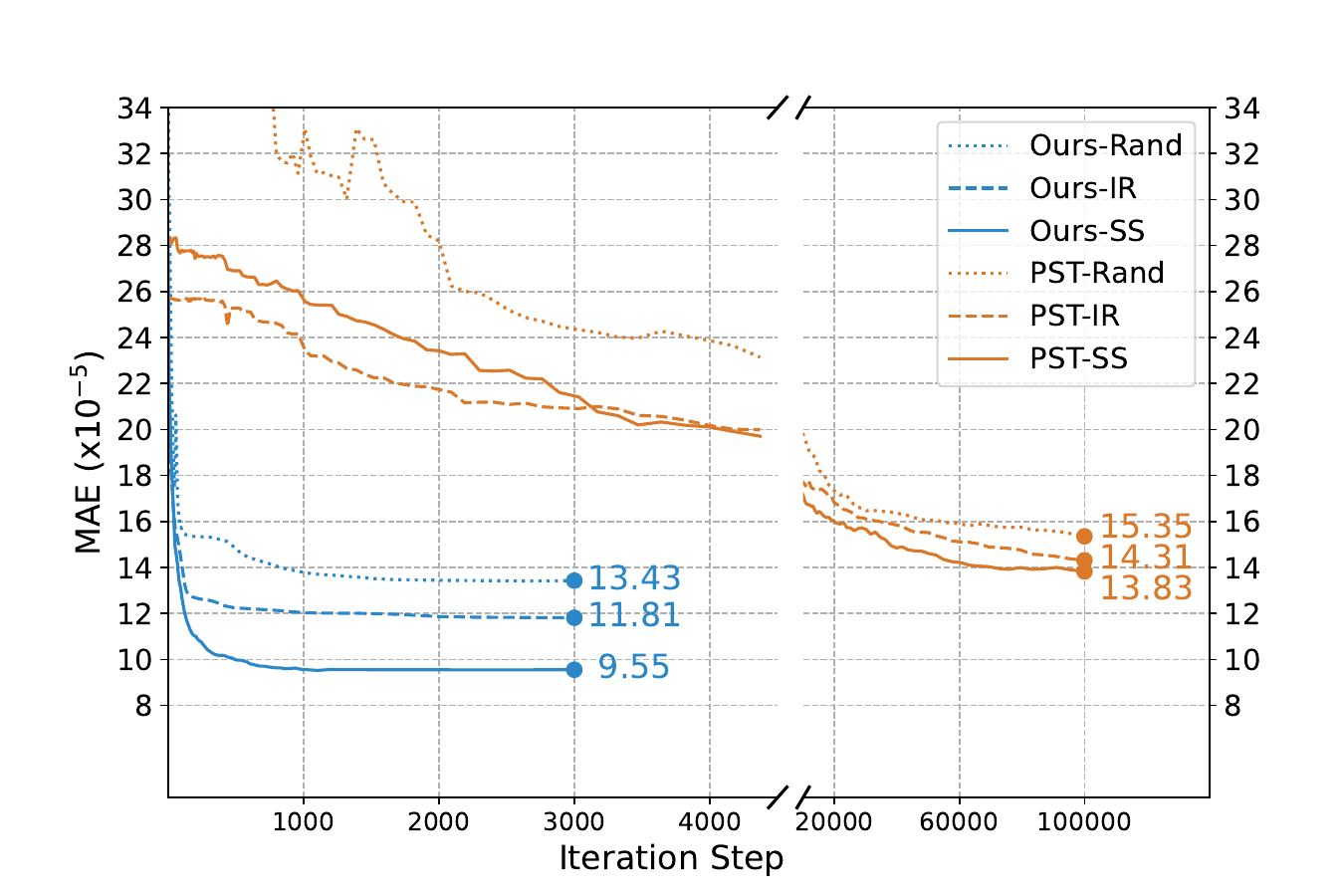}
    \end{minipage}
    \hfill 
    \begin{minipage}{0.495\columnwidth} 
        \centering
        \includegraphics[width=\linewidth]{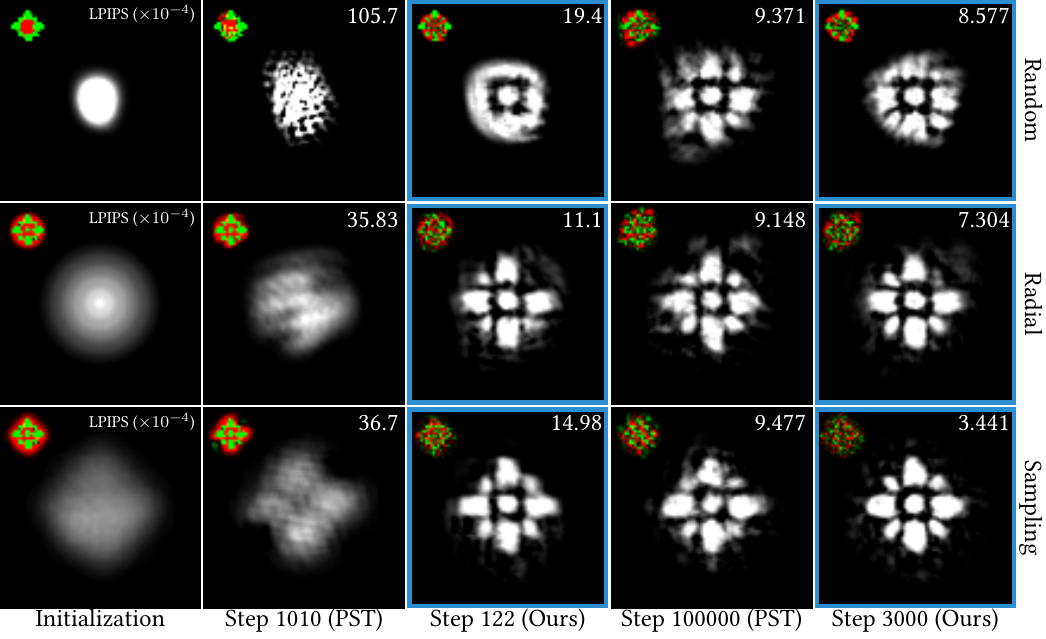}
    \end{minipage}
    \centering
    \begin{minipage}{0.495\columnwidth} 
        \centering
        \includegraphics[width=\linewidth]{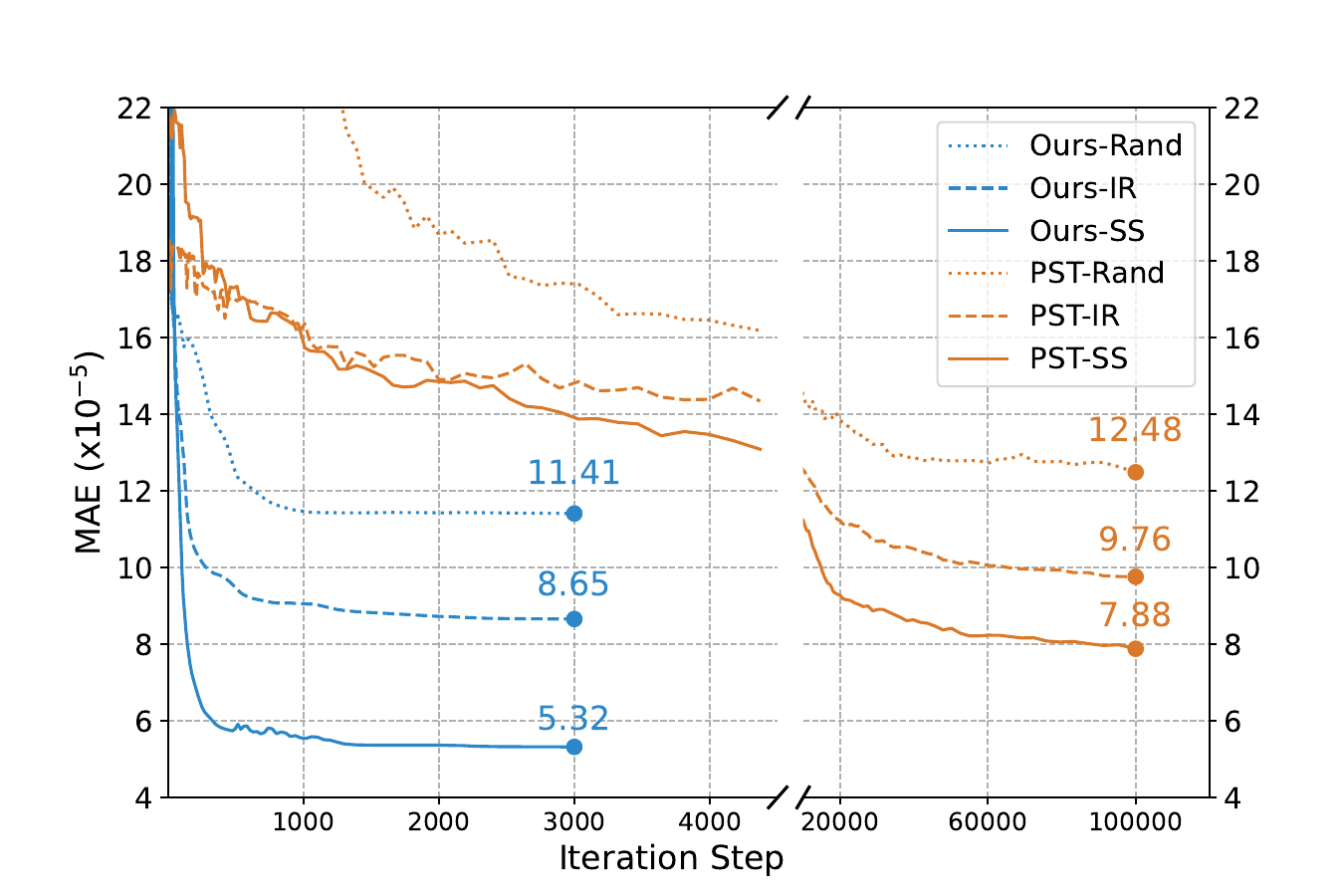}
    \end{minipage}
    \hfill 
    \begin{minipage}{0.495\columnwidth} 
        \centering
        \includegraphics[width=\linewidth]{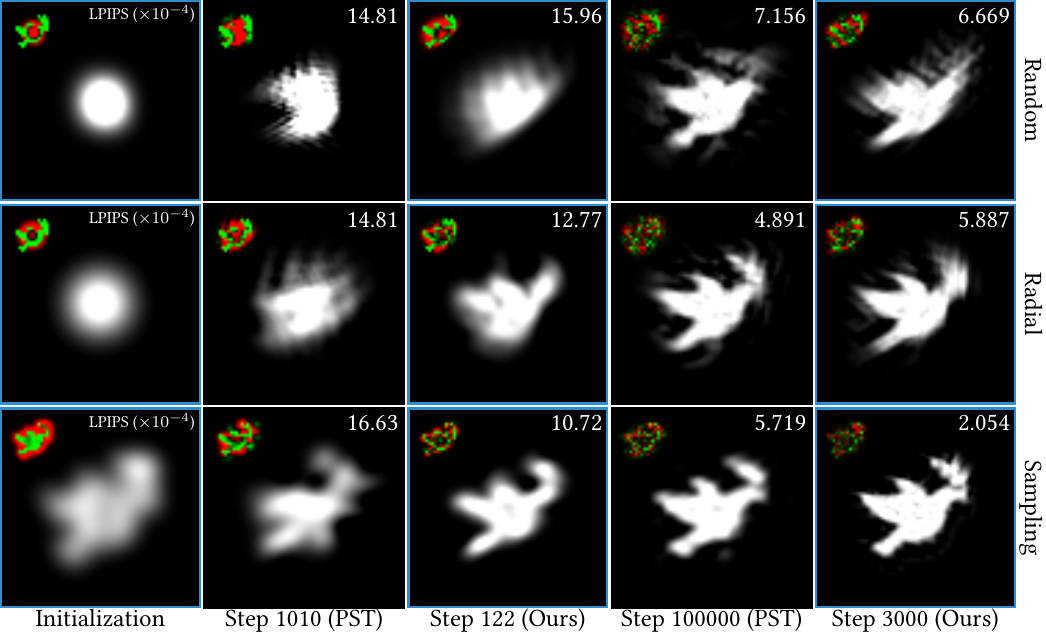}
    \end{minipage}
    \caption{\textbf{Ablation of initialization strategies on the \emph{Flower} and \emph{Dove} kernel.} \chg{We evaluate both our method and Parallel Simulated Annealing (PST) combined with three initialization schemes: Random (Rand), Increasing Radial (IR), and Sparse Sampling (SS).}}
    \label{fig:init-abl-flower}
\end{figure}
\begin{figure}[t]
    \centering
    \begin{minipage}{0.38\columnwidth} 
        \centering
        \vspace{-2pt}
        \includegraphics[width=\linewidth]{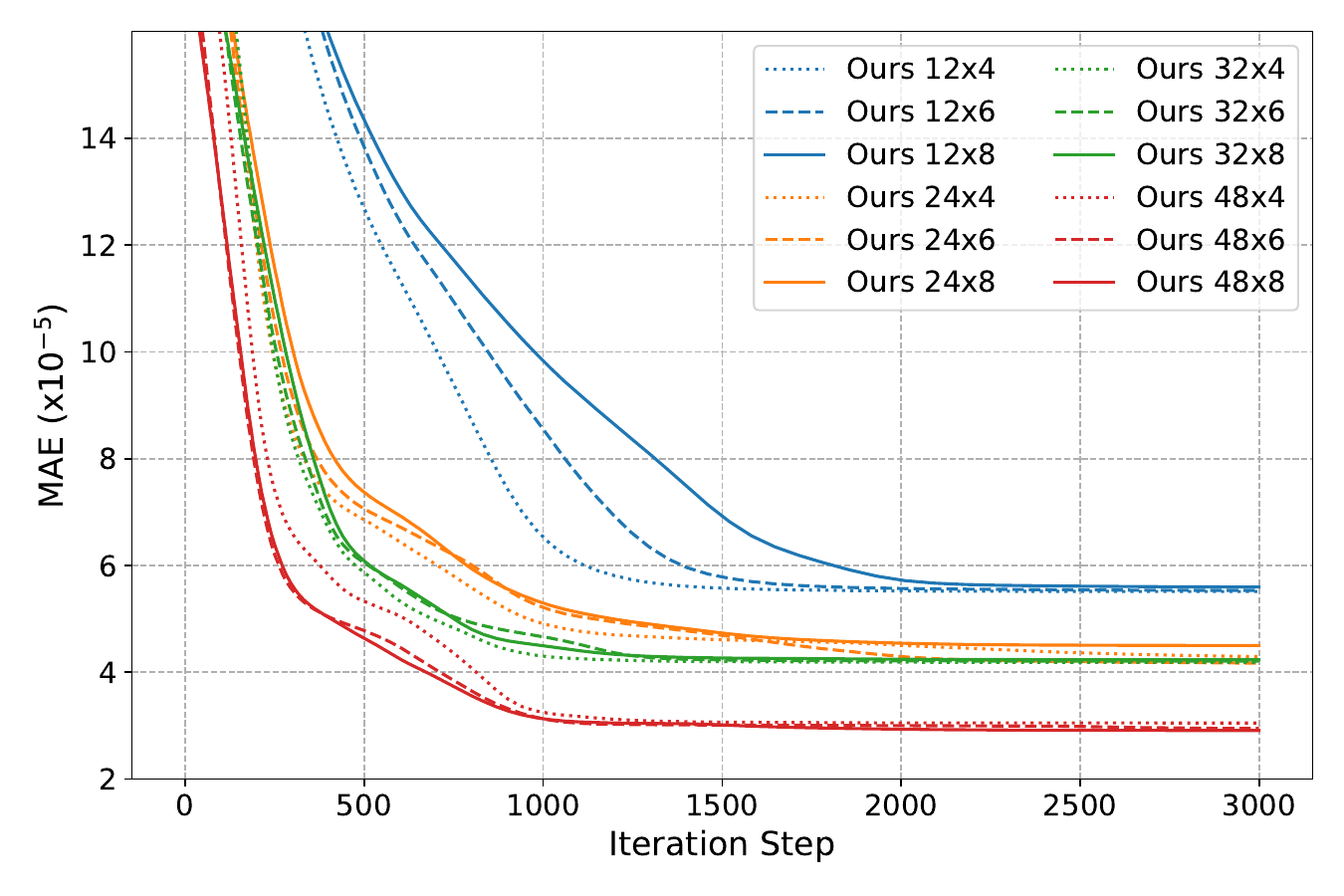}
    \end{minipage}
    \hfill 
    \begin{minipage}{0.3\columnwidth} 
        \centering
        \vspace{-8pt}
        \includegraphics[width=\linewidth]{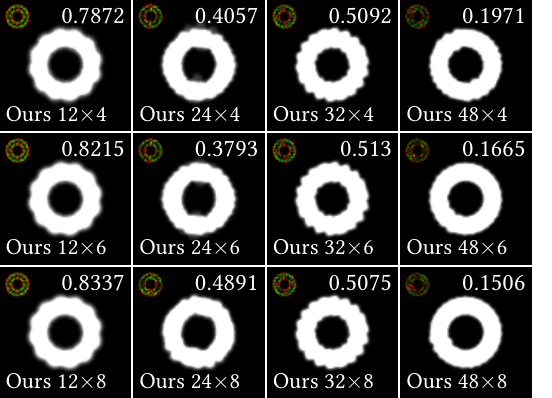}
    \end{minipage}
    \begin{minipage}{0.3\columnwidth} 
        \centering
        \vspace{-8pt}
        \includegraphics[width=\linewidth]{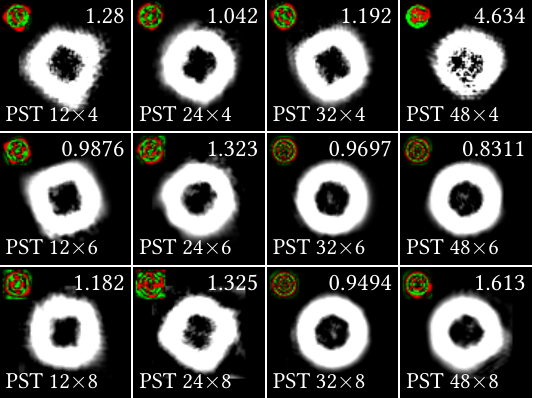}
    \end{minipage}
    \caption{\textbf{Ablation results for various configurations of samples and layers on Ring kernel.}}
    \label{fig:layer-abl-ring}
\end{figure}
We conduct ablation studies to validate our main design choices, focusing on both initialization strategies and different layer configurations.

We first evaluate different initialization schemes across multiple kernels, as shown in~\refFig{fig:init-abl-flower}. Both our method and Parallel Simulated Tempering (PST) benefit from the proposed Sparse Sampling (SS) initialization, which consistently outperforms the Increasing Radial (IR) initialization, while the Random (Rand) initialization performs worst. Although SS accelerates convergence for both our method and PST, PST still requires more than 30× the number of iterations to converge compared with ours, and our final reconstruction quality is significantly higher.

We further study the influence of different configurations, varying the number of layers and the number of samples, as shown in~\refFig{fig:layer-abl-ring}. The convergence curves show that all configurations converge stably, and configurations with more samples and layers tend to achieve higher quality. Compared with PST, our method delivers more consistent behavior and better quality across all tested configurations.

For additional results, please refer to the Appendix, which includes ablations on Gaussian kernels with fewer samples and quantitative evaluations of initialization and regularization strategies on arbitrary kernels.

\section{Discussion and Conclusion}
We introduced a differentiable framework that recasts the challenging problem of approximating large, complex convolution kernels as an end-to-end optimization task. Our approach supports a wide range of kernels—from simple Gaussians to complex, non-convex forms—and converges to high-fidelity solutions far more efficiently than prior methods. We extend this with filter-space interpolation, enabling multi-dimensional spatially varying effects with minimal per-pixel overhead. A current constraint of our formulation is that it requires access to the target dense kernel during optimization and filtering.
This work opens several promising avenues for future research, including multi-dimensional parameter maps for simultaneous control over kernel attributes and the use of neural architecture search to discover hardware-optimized filter decompositions. Overall, our method provides a practical, high-performance solution for advanced image filtering in real-time applications such as computational photography, while remaining fully differentiable and thus usable as a trainable layer within modern deep learning pipelines.
\section*{Acknowledgments}
This work was partially supported by National Key R\&D Program of China (No. 2024YFB2809104), NSFC (No. 52532013), and Key R\&D Program of Zhejiang (No: 2025C01064). We also thank the anonymous reviewers for their constructive comments.

\bibliography{reference}
\bibliographystyle{style/reference}

\end{document}